\documentclass[aps,prb,a4paper,groupedaddress, amsfonts, amssymb, amsmath,twocolumn,10pt,showkeys,nolongbibliography]{revtex4-2}
\usepackage[english]{babel}

\usepackage{hyperref}
\usepackage{graphicx}
\usepackage{color}
\usepackage[normalem]{ulem}

\begin{document}

\title{Half-integer quantum anomalous Hall effect in MnBi$_2$Te$_4$ \\ 
via non-magnetic extension}

\author{Igor P. Rusinov}
\affiliation
{Lebedev Physical Institute of the Russian Academy of Sciences, Moscow, Russia}
\affiliation{St. Petersburg State University, 199034  St. Petersburg, Russia}
\email{rusinovip@gmail.com}
\author{Tatiana V. Menshchikova}
\affiliation{Lebedev Physical Institute of the Russian Academy of Sciences, Moscow, Russia}
\affiliation{St. Petersburg State University, 199034  St. Petersburg, Russia}
\author{Evgueni V. Chulkov}
\affiliation{Donostia International Physics Center (DIPC), 20018 San Sebasti$\acute{a}$n-Donostia, Euskadi, Spain}
\affiliation{St. Petersburg State University, 199034  St. Petersburg, Russia}

\date{\today}

\begin{abstract}

Within an ab-initio approach, we study the electronic structure and Hall response of the antiferromagnetic topological insulator film of MnBi$_2$Te$_4$ in contact with non-magnetic GeBi$_2$Te$_4$ (Bi$_2$Te$_3$) overlayer. The proposed junction design eliminates the exchange splitting one of the Dirac cones of the MnBi$_2$Te$_4$ (0001) thus manifesting a half-integer quantum anomalous Hall phase. The described recipe allows to enlarge temperature of the effect observation, enriches the diversity of topological properties of the MnBi$_2$Te$_4$ compound and is appropriate for other layered antiferromagnetic topological insulators. Applying the proposed heterostructure design, the half-quantized Hall phase can be established with employing other overlayer and magnetic topological insulator films.
\end{abstract}


\maketitle

\section{Introduction}

Materials combining spin–orbit and magnetic interactions attract research interest for the study of new effects arising from the interplay of band topology and magnetism, as well as for their potential applications in spintronic devices \cite{HIROHATA2020166711, Yan2024}.
Among the widely studied materials, the class of van der Waals magnetic topological insulators (TIs) based on the MnBi$_2$Te$_4$ septuple layer blocks and related compounds stands out, having proven to be a versatile platform for such investigations~\cite{PhysRevB.81.245209, menshov2019formationmagnetic,Vyazovskaya2025}. Indeed, a wide range of phenomena has been proposed and experimentally confirmed in these materials: the formation of the antiferromagnetic TI phase~\cite{otrokov2019predictionobservation}, the quantum anomalous Hall effect (QAH)~\cite{otrokov2019uniquethicknessdependent,deng2020quantumanomalous} and M\"{o}bius insulating phase~\cite{PhysRevLett.124.136407}.

Owing to magnetic order and preserving combination of time-reversal symmetry with half-translation ($\hat T \hat \tau_{1/2}$), the remarkable characteristic of MnBi$_2$Te$_4$ is axion insulating phase~\cite{zhang2019topologicalaxion,liu2020robustaxion}. It results in the emergence of unique magnetoelectric and optic phenomena and formation of chiral low-dimensional topological states. At vacuum boundaries, for instance, both exchange-split and gapless topological surface states (TSS) can form, depending on the specific surface termination~\cite{otrokov2019predictionobservation, zhang2019topologicalaxion}. Additionally, low-dimensional states arise on crystal edges as a consequence of higher-order topology~\cite{PhysRevLett.124.136407, Gu2021, Yan2024}. An important property of the aforementioned topological states lies in their formation related to/with the magnetic configuration, which allows quantum transport manipulation.



The realization of an axion insulating phase enables the probing of the half-integer QAH ($\frac{1}{2}$-QAH) effect (or, alternatively, the parity anomaly~\cite{PhysRevLett.53.2449, PhysRevLett.61.2015}) in magnetic topological systems, which represents a significant scientific challenge. This effect is characterized by the formation of a single split Dirac state at the surface, which gives rise to a fractional Hall conductivity of $\sigma_{xy}=\frac{e^2}{2h}$.
In addition, the longitudinal conductivity $\sigma_{xx}$ is finite, which distinguishes the $\frac{1}{2}$-QAH state from the semiconducting character inherent to the QAH state. The $\frac{1}{2}$-QAH effect has attracted significant attention as a platform where condensed matter physics merges with quantum field theory~\cite{Dudal2022, PhysRevB.94.195150}, thereby opening a route toward the possible detection of dark matter. This effect is also closely linked to the physics of Majorana fermions, which opens up a pathway toward practical quantum computation.

Experimental verification of the $\frac{1}{2}$-QAH effect has been demonstrated in 'semi-magnetic' TIs: Cr-doped (Bi/Sb)$_2$Se$_3$~\cite{Mogi2022} at 2K, and  V- and Cr-doped (Bi/Sb)$_2$Te$_3$~\cite{vxcb-rwbl} films.
Another kind of the system has been fabricated employing magnetic overlayer~--- 
(Bi,Sb)$_2$Se$_3$/Cr$_2$Ge$_2$Te$_6$~\cite{jain2024} allows the temperature to be improved up to $\sim$13~K.
In the presented examples, a single Dirac cone is gapped for the surface with the overlayer, which impacts optical and transport measurements~\cite{Mogi2022,jain2024}, whereas the pristine surface is characterized by a gapless spectrum.

\begin{figure*}[t]
        \centering
        \includegraphics[width=\textwidth]{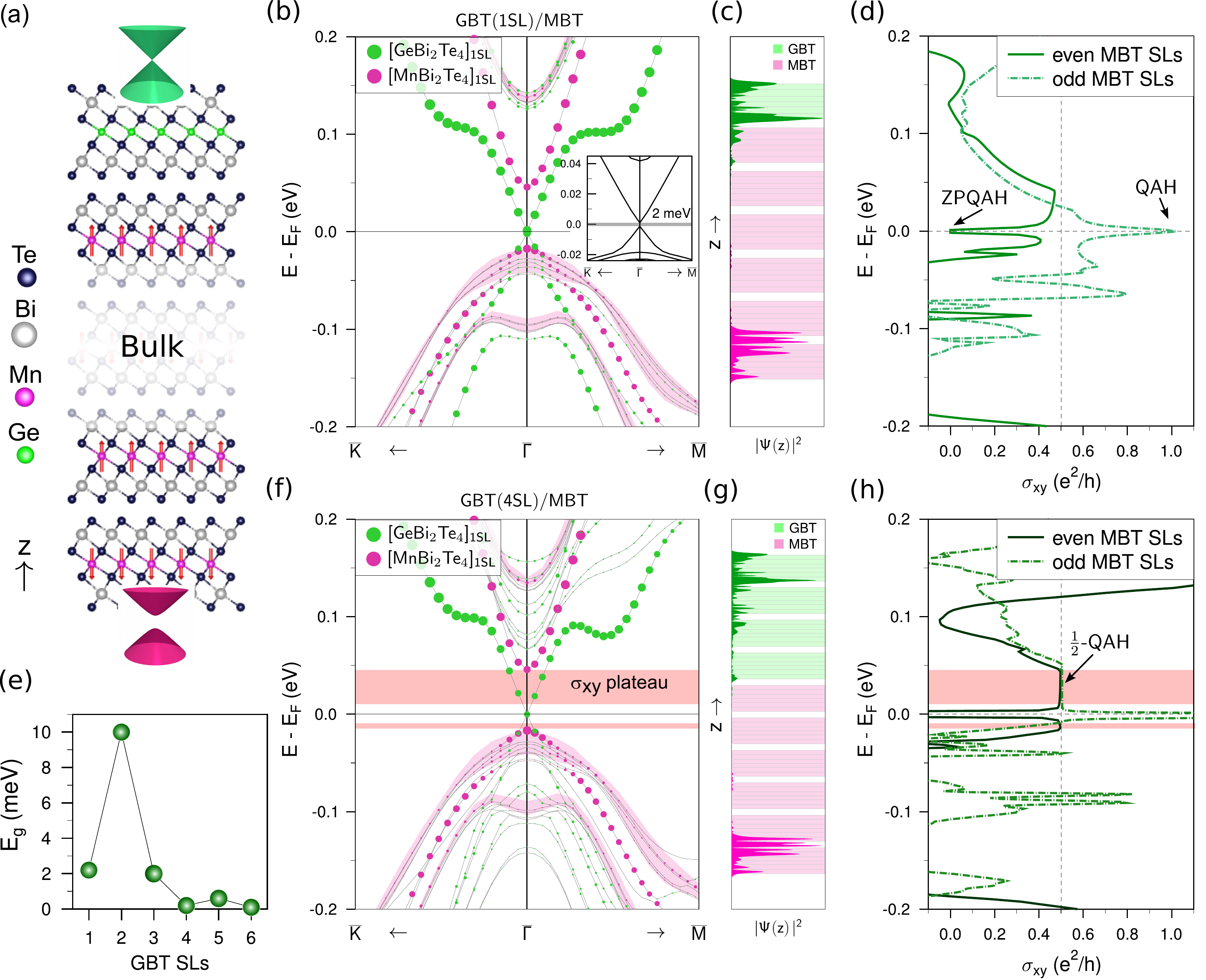}
        \caption{\label{fig1} (a) Schematic illustration of crystal structure of GBT(1SL)/MBT and Dirac cones for both terminations. (b,f) Surface band structure of GBT(1SL)/MBT and GBT(4SL)/MBT, respectively. The size of magenta (green) 
        circles reflects the weight of the state in outer SL block for lower (upper) termination.  
        The inset in the panel (a) represents the momentum-energy area of the Dirac cone with tiny exchange splitting for GBT termination. The energy range of half-quantized Hall plateau is denoted by light red areas in panel (f). (c,g) Charge density
distribution integrated over xy plane at the $\bar\Gamma$ point for the surface states located at the global bulk band gap. (e) The magnetic exchange gap width with respect to the number of SL-blocks, $N$, for GBT($N$SL)/MBT heterostructure. (d,h) Hall conductivity for GBT(1SL)/MBT and GBT(4SL)/MBT, respectively. The energy ranges corresponding to QAH, zero-plateau QAH (ZPQAH), and $\frac{1}{2}$-QAH are indicated on the panels by arrows.}
\end{figure*}
 
In the present study, we propose and analyze an alternative approach of introducing the $\frac{1}{2}$-QAH state. It consists in forming heterostructures composed of a MnBi$_2$Te$_4$ film and a non-magnetic semiconductor overlayer of GeBi$_2$Te$_4$ or Bi$_2$Te$_3$.
We demonstrate that the non‑magnetic side of the heterostructure hosts a TSS with a tiny exchange gap, even after the deposition of just a few van der Waals blocks. As a result, the topological transport is governed by the exchange gap of the magnetic side of the heterostructure, which leads to half‑quantized net conductivity via the proposed non‑magnetic deposition.

\section{Results}

The key ingredient of the heterostructure is a multilayer film of the antiferromagnetic TI MnBi$_2$Te$_4$ (MBT), characterized by the required physical properties: an out-of-plane easy axis and hosted exchange split TSS~\cite{otrokov2019predictionobservation,zhang2019topologicalaxion}.
Referring to these states as “topological” in what follows is justified by the bulk band topology of MnBi$_2$Te$_4$; this terminology is supported by their robustness against severe surface degradation~\cite{PhysRevX.9.041038}, despite the absence of direct symmetry protection for them. The non‑magnetic semiconductor component comprises ultrathin films of the septuple‑layer (SL) GeBi$_2$Te$_4$ (GBT) or the quintuple‑layer (QL) Bi$_2$Te$_3$ (BT).  The selection of this component is governed by two key criteria: favorable in-plane lattice matching and appropriate work function alignment~\cite{NanoLett2013Mensh,JetpLett2013Mensh,Menshov_2014}. Also, the number of MBT SL blocks is chosen to be sufficient to exclude interaction between MBT surface states (6SL and 7SL blocks). An illustration of "semi-magnetic" heterostructures under study is shown in Figure~\ref{fig1}.a, for the case of GBT(1SL)/MBT. Thus, the considered systems are asymmetric, and the two 
MBT surfaces are non-equivalent forming pristine and interface sides.

First, consider the GBT(1SL)/MBT junction electronic structure (Figure \ref{fig1}.b). It features two spin‑split Dirac states located on opposite sides of the heterostructure. The first one resides at the MBT pristine surface and exhibits an exchange splitting of $\sim$60~meV at the $\overline{\Gamma}$ point, consistent with previous results~\cite{SHIKIN2023414443}. For the interface side, the splitting is reduced to $\sim$2~meV, which results from the redistribution of the TSS into the nonmagnetic GBT 1SL block. The remaining gap originates both from the state's tail within the interface MBT layer (Figure \ref{fig1}.c) and from the induced magnetization of the surface GBT block. In contrast, the pristine side of the heterostructure remains gapped and is unaffected by the non‑magnetic deposition on the opposite side, due to the absence of interaction between them. The exchange‑split Dirac state of the pristine surface side is localized mostly near the van der Waals gap of the outer MBT SLs. The presented picture of electronic structure agrees with earlier report.~\cite{nano13142151}

Essentially, the spectrum under study lacks additional interface states at the Fermi level \cite{NanoLett2013Mensh,JetpLett2013Mensh,Menshov_2014}, which is attributed to the specific alignment of the work functions between the magnetic and non‑magnetic parts. Indeed, the isolated constituents of the GBT(1SL)/MBT heterostructure exhibit semiconducting behavior: the MBT gap lies nearly within the wide semiconducting gap of GBT(1SL) (Supplemental materials (SM) sec.I, Figure S1.a). Note, the resulting spectrum on the non‑magnetic side closely resembles that observed on the surface of a GBT 3D TI.~\cite{Menshchikova2011,nano13142151}

As can be seen in Figure~\ref{fig1}.d, within a narrow region of the tiny exchange gap at the interface side the corresponding Hall conductivity, $\sigma_{xy}$, exhibits behavior consistent with an axion insulating state of free‑standing MBT~\cite{Zhuo2023}: $\sigma_{xy}=0$ [zero-plateau QAH] and $e^2/h$ [QAH] for an even and odd number of magnetic SL blocks, respectively.
The magnetic part of the employed 6SL MnBi$_2$Te$_4$ film is itself correctly described within the axion picture, since the two TSSs are essentially decoupled and bulk-like states are formed; it is also considered an axion insulator in both theoretical~\cite{PhysRevResearch.4.L022067} and experimental~\cite{liu2020robustaxion,Lin2022} [Lin et. al Nature Comm. 13, 7714 (2022)] investigations. In our calculations, the deviation from quantization is tiny and corresponds to an axion angle of $\theta=1.04\pi$, which is a residual finite-size effect.

Outside the tiny plateau of quantization energy window and within the pristine surface exchange gap, the $\sigma_{xy}$  approaches the half‑quantized value of $0.5~e^2/h$, without formation the plateau behavior. The deviation from the half-quantized value arises from the reduced  $\sigma_{xy}$ contribution associated with the interface side of the heterostructure. The described reduction is independent of the magnetic moment direction of the interface MBT block, which is opposite for odd and even numbers of MBT SLs. However, the retaining contribution  is still sufficient owing to the magnetic proximity effect~\cite{PhysRevB.88.224401, PhysRevB.88.144430}, as the TSS is partially localized within the interface MBT layer (Figure~\ref{fig1}.c); hence, the $\frac{1}{2}$-QAH state is not induced in this regime.

Upon increasing the number of GBT overlayer blocks, the exchange gap displays an oscillatory behavior (Figure~\ref{fig1}.e), vanishing at 4–6 SLs, while the spectra near the Fermi level do not show any dramatic modifications compared to the GBT(1SL)/MBT case.
On the other hand, TSS becomes localized in the spatial region where magnetic proximity is suppressed (see SM, sec. II, Figure S2). Thus, in the GBT(4SL)/MBT case (Figure~\ref{fig1}.f,g), a gapless, topologically protected Dirac cone state is hosted mainly within the two outer GBT SLs (Figure~\ref{fig1}.g).  Owing to this,  the system  exhibits $\frac{1}{2}$-QAH phase over the wide energy range (Figure~\ref{fig1}.h), excluding the region of infinitesimal exchange gap. The described deviation arises owing to the retained weak magnetic proximity effect (Figure~\ref{fig1}.g), vanishing over further increasing the number of non‑magnetic blocks, and is caused by non-zero temperature broadening parameter for Fermi-function employing at Hall conductivity calculations (see SM sec. III, Figure S3). The presented picture of $\frac{1}{2}$-QAH is unaffected by magnetic moment direction of interface MBT inducing both for odd and even number of MBT SLs. 

\begin{figure*}[t]
\centering
\includegraphics[width=0.7\textwidth]{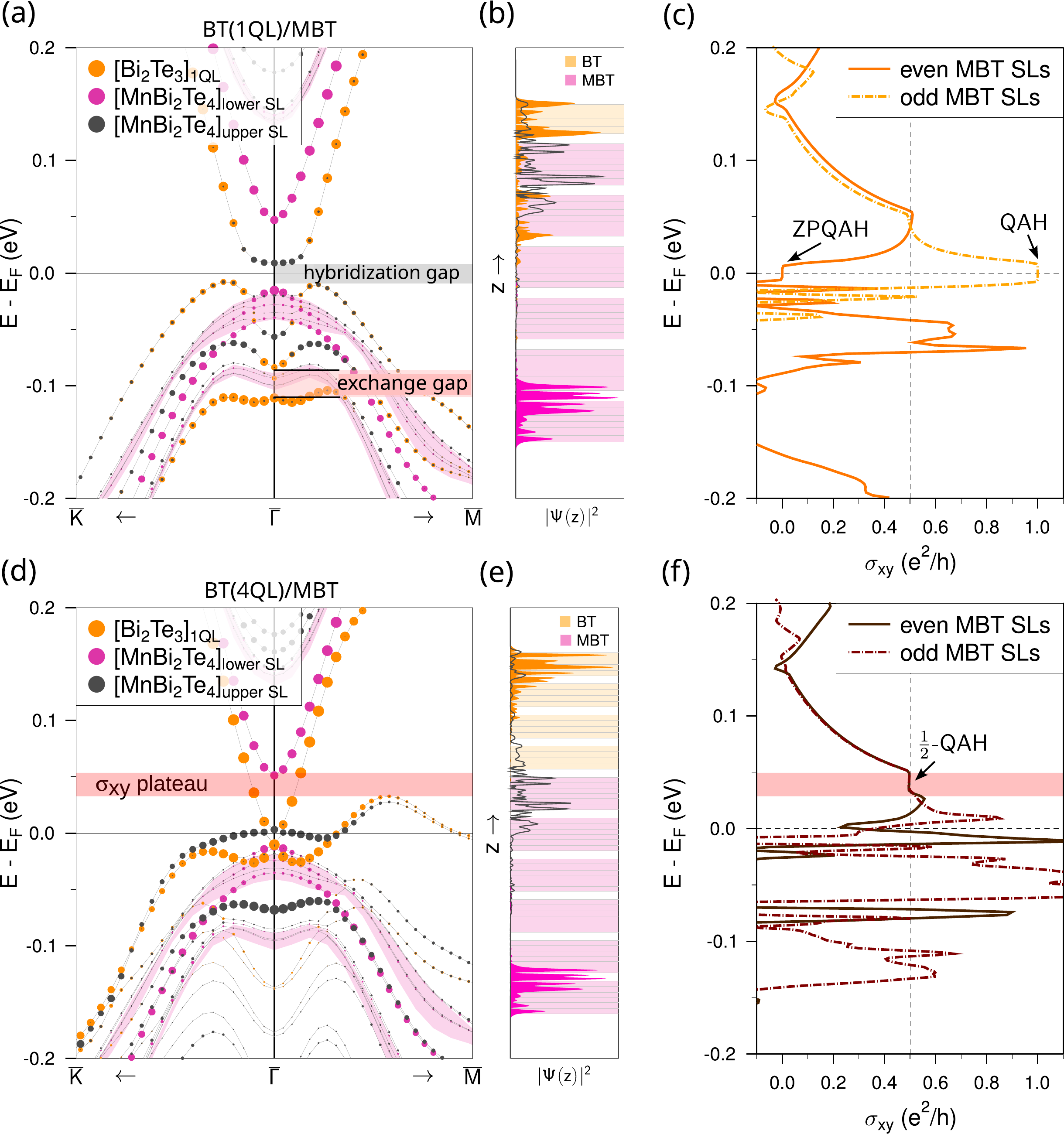}
\caption{\label{fig2}
        (a,d) Electronic spectra of BT(1QL)/MBT and BT(4QL)/MBT, respectively. The radia of colored circles are denoted the wight of TSS in the outer structural block. The energy range of half-quantized Hall plateau is denoted by light red areas in panels (d,f). (b,e) Charge density distribution integrated over xy plane at the $\bar\Gamma$ point for the surface states located at the global bulk band gap: TSS in orange and magenta, trivial surface state in gray for BT(1QL)/MBT and BT(4QL)/MBT, respectively. (c,f) Hall conductivity for BT(1QL)/MBT and BT(4QL)/MBT, respectively.}  
\end{figure*}

\begin{figure*}[t]
\includegraphics[width=\textwidth]{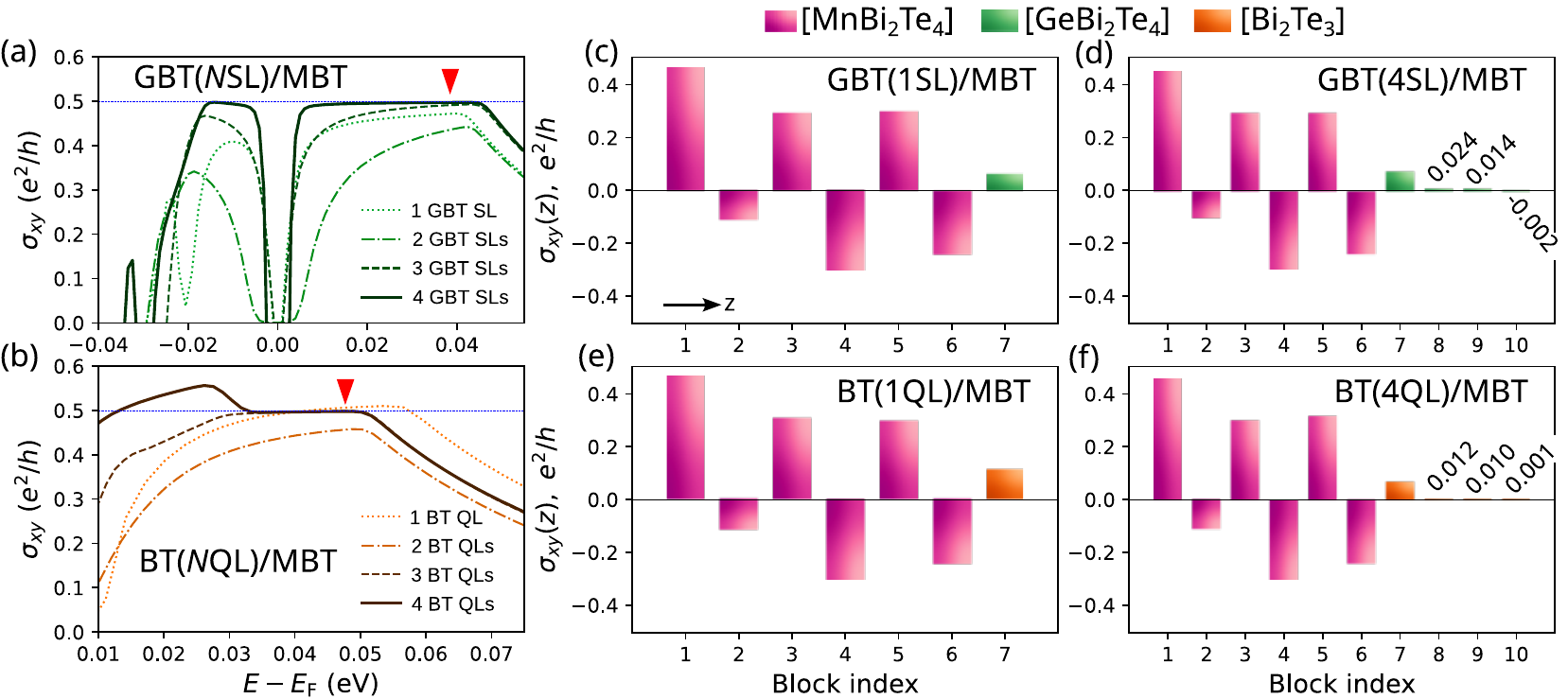}
    \caption{\label{fig3}
    Hall conductivity  near the plateau energy range for different number of overlayer blocks and 6SLs of MBT blocks. (c-f) Spatial distribution of $\sigma_{xy}(z)$ for (c) GBT(1SL)/MBT, (d) GBT(4SL)/MBT, (e) BT(1QL)/MBT and (f) BT(4QL)/MBT heterostructures, for the energy values denoted by triangles in panels (a) and (b).}
\end{figure*}

As shown in Figure~\ref{fig1}.d, the $\frac{1}{2}$-QAH effect is also accompanied by the presence of metallic character of band structure, since the interface side Dirac cone crosses the band gap. 
As a result, a nonzero longitudinal conductivity $\sigma_{xx}$ does not vanish. Indeed, experimental measurements Refs.~\cite{Mogi2022, Wang_parity_2026, vxcb-rwbl}, demonstrate the minimum value of the longitudinal conductivity in magnetically doped (Bi,Sb)$_2$Te$_3$ heterostructures of $\sim$0.6 e$^2$/h, which is consistent with the  theoretical calculations of~\cite{ Wang_parity_2026}. Thus, it can be concluded that the value of $\sigma_{xx}$ in the proposed system itself depends on the position of the Fermi level, temperature, and crystal structure imperfections.

To probe the $\frac{1}{2}$-QAH phase stability against heterostructure design, we turn to the BT/MBT junction (Figure \ref{fig2}). For the a case of single QL overlayer (Figure \ref{fig2}.a,b), the TSS located on the non-magnetic side exhibits a finite exchange gap of $\sim$27~meV and energetically hosts in the occupied part of the spectrum, at $\sim$~-0.1~eV. The gap is wider than in the GBT(1SL)/MBT case, owing to the TSS being localized within both the BT overlayer (subject to magnetic proximity) and the two interface MBT blocks (Figure \ref{fig2}.b).

The remarkable feature of the spectrum is additional trivial interface quantum well states inherited from the projection of bulk bands (black circles in the figure). One of them appears at the Fermi level, near the TSS, and both states spatially overlap, being confined to the two outermost MBT blocks and the non-magnetic BT block (Figure~\ref{fig2}.b). As a result, they form hybridization band gap which does not depend on magnetic orientation in the system (SM, sec. IV, Figure S4). Note, similar interface states and hybridization gap were also observed for surface electronic structure of MnBi$_2$Te$_4$$\cdot$(Bi$_2$Te$_3$)$_n$~\cite{Klimovskikh2020, Sitnicka_2022} and related superlattices~\cite{PhysRevB.110.205113}. 
The BT/MBT heterostructure exhibits a more complex electronic structure due to the specific work‑function alignment (SM, sec. I, Figure S1.c), wherein the band gaps of the heterostructure constituents do not overlap and the gap of the non‑magnetic overlayer lies below that of the magnetic layer~\cite{NanoLett2013Mensh,JetpLett2013Mensh,Menshov_2014}.

As shown in Figure~\ref{fig2}.c, the Hall conductivity is nonzero and approaches half-quantized value within the energy range of 30--60 meV above the Fermi level. However, the characteristic plateau is absent in this case, owing to the combined contributions of the junction-side TSS and the interface quantum well state. At the same time, the 16~meV hybridization gap acts as an exchange gap, thereby introducing axion insulating plateau behavior within the corresponding energy range, with a well‑defined QAH (zero‑plateau QAH) at odd (even) numbers of MBT SLs.
 
As the BT overlayer thickness is increased to 4 QL (Figures~\ref{fig2}.d and~\ref{fig2}.e), the TSS on the interface side shifts upward toward the Fermi level and becomes localized within the two outer non-magnetic QL blocks, leading to the vanishing of the exchange gap. Simultaneously, quantum well states give rise to a clearly visible electron pocket along the $\overline{\Gamma}$-$\overline{M}$ direction.
Their spatial localization pattern is similar to that observed in the BT(1QL)/MBT case. However, in the present case, the two distinct types of surface states are spatially separated, and no hybridization gap emerges. The calculated $\sigma_{xy}$ displays a well‑defined half‑quantized plateau, thus demonstrating the universality of the proposed approach. Nevertheless, the corresponding energy window is narrower than that in the GBT(4SL)/MBT system, due to the contribution from interface states. This contribution remains unchanged, as the localization of these states does not vary with an increasing number of BT blocks and is determined solely by the interface region. As before, deviations from the $\sigma_{xy}$ plateau appear near the small exchange gap.

Described differences in the surface electronic structures between BT/MBT and GBT/MBT heterostructures, arising from the corresponding work function alignment of the constituents, manifest in the variation of $\sigma_{xy}$  upon increasing the thickness of the non‑magnetic overlayer, which is illustrated in Figure~\ref{fig3}.a,b for the energy range near $\frac{1}{2}$‑QAH plateau. First, $\frac{1}{2}$-QAH emerges more monotonically for the GBT/MBT, owing to the absence of additional interface states. As for the BT/MBT, the formation of the plateau is achieved already at 3QL overlayer, while for GBT/MBT~--- at 4SL, which is provided by stronger localization of interface side TSS for the latter. It is also reflected in the spatial resolution of $\sigma_{xy}(z)$ (Figure~\ref{fig3}.c-f), which is demonstrated for the energy denoted by triangles in Figure~\ref{fig3}.a,b. The contribution into  $\sigma_{xy}(z)$ involves MBT film with the tail inside the non-magnetic overlayer part. Thus, with increasing thickness of the non‑magnetic layer, the contribution becomes more distinctly separated from the TSS formation region. From the figures is evident, in the thin GBT(1SL)/MBT (BT(1QL)/MBT) overlayers $\sigma_{xy}$ defines for the entire film (Figures~\ref{fig3}.c,e), while for the thick GBT(4SL)/MBT (BT(4QL)/MBT) it takes place inside MBT with the tail into non-magnetic block (Figures~\ref{fig3}.d,f) unaffecting the spatial region of three SL (QL), where TSS is localised. The existence of the $\sigma_{xy}$  contribution for closer to MBT non-magnetic block is caused by magnetic proximity effect. The remarkable characteristic of $\sigma_{xy}(z)$ is its oscillation inside MBT for all the considered heterostructures, which is originated from the axion insulating phase of this compound and previously reported in Refs.~\cite{varnava2018surfacesaxion} and \cite{Gu2021}.

\section{Discussion}

The presented heterojunctions exhibiting $\frac{1}{2}$-QAH effect have key structural differences compared to previously studied within theory~\cite{otrokov2017highlyorderedwide, Petrov2019, PhysRevMaterials.6.074205, Muzaffar_2025, Muzaffar_2025_2, sattar2026} and experiment~\cite{Mogi2022, vxcb-rwbl, jain2024, Wang_parity_2026}.  In those systems thin magnetic films or magnetic impurities are deposited on a non-magnetic TI substrate or thick film. 
Hence, the overlayer thickness increasing leads to both the exchange gap width (because the parity of the blocks changes) and its position relative to the Fermi level (because the work function of the films depends on the thickness) change. In our approach, these quantities are fixed and do not depend on the thickness of the deposited blocks.

In principle, the two-dimension nature of the magnetic layer tends to restricts the temperature range over which the half quantization can be observed. However, it has been found that surface deposition of magnetic SL blocks allows to improve critical temperature owing to the formation of additional potential gradient near the surface magnetic region. 
Thus, theoretical studies of heterostructures composed of a monolayer (bilayer) of MnBi$_2$Te$_4$ on top of a TI have reported $T_C\sim$20~K ($T_N\sim$38.7~K), attributed to enhanced magnetic anisotropy~\cite{Muzaffar_2025, Muzaffar_2025_2}. In the present work, the critical temperature corresponds to the bulk N\'{e}el temperature of MnBi$_2$Te$_4$, $T_N\sim$24~K.
In such a way, the use of related  compounds, such as MnSb$_2$Te$_4$~\cite{Wimmer_2021, Kochetkova2025} or V-based magnetic TIs~\cite{PhysRevB.103.235142}, offers promising avenues for enhancing the temperature range up to $\sim$40--90~K.

It should be noted that the critical temperature sets the upper limit for observing the $\frac{1}{2}$-QAH effect. This limit is further reduced by structural and magnetic imperfections, which determine the energy gap and the position of the Fermi level, and thus govern the thermal activation of bulk charge carriers that degrades the quantization. The actual and realistic estimation is of 1-2~K, corresponding to the case observation of QAH and ZPQAH in thick films of MnBi$_2$Te$_4$~\cite{deng2020quantumanomalous, liu2020robustaxion}.

Additionally, the proposed design implies the possibility the realization of the $\frac{1}{2}$-QAH by depositing magnetic thick films onto non-magnetic TI substrates, such as chosen in this study Bi$_2$Te$_3$ or PbBi$_2$Te$_4$. The mentioned configuration corresponds to the case BT($N$QL)/MBT and GBT($N$SL)/MBT in the limit of $N\to\infty$. Furthermore, the Fermi level, and the N\'eel temperature can be tuned via substitution with Ge/Pb/Sn, Bi/Sb or Te/Se~\cite{Chen_doped2019, PhysRevB.100.104409}, or via additional doping~\cite{Du2021}. To access the exchange gap windows in which the $\frac{1}{2}$-QAH is activated, the Fermi level can be further fine-tuned in transport experiments via gate voltage.

Another way involves the surfaces of "semi-magnetic" superlattices (MnBi$_2$Te$_4$)$\cdot$(Bi$_2$Te$_3$)$_n$~\cite{Klimovskikh2020, PhysRevX.11.011039, Han2026} terminated by
non-magnetic Bi$_2$Te$_3$ structural blocks. However, this approach faces challenges due to  a reduced magnetic critical temperature down to $\sim$10~K~\cite{Yao2024}. Such reduction arises from weakened interlayer magnetic interactions, and the formation of additional interface states similar to those observed into the BT/MBT heterostructure described in the present investigation.

The scenario discussed for BT($N$QL)/MBT and GBT($N$SL)/MBT ($N<4$) [when one of the exchange gaps width and corresponded Hall conductivity are suppressed] can also be realized by depositing topologically trivial BiTeI on top of MBT at the Te-Te side contact junction~\cite{PhysRevB.107.045402}. However,
topolgically-trivial nature of BiTeI restricts the formation of $\frac{1}{2}$-QAH.
Also, the same effect can be engineered by modulating surface electrostatic potential ~\cite{PhysRevB.104.115168} by increasing surface van der Waals distance~\cite{SHIKIN2023414443} or introducing controlled surface defects~\cite{PhysRevB.104.115168,PhysRevLett.130.126702,Ahn2023,Yang_2025}.


\section{Conclusion}
In summary, our ab initio study demonstrates a generic and effective route to realizing the half-quantized quantum anomalous Hall effect by interfacing an antiferromagnetic topological insulator with a non-magnetic overlayer. This interface engineering strategy allows not only overcomes the conventional temperature limitations but also enriches the functional diversity of the material class. Using the MBT film and non-magnetic GBT or BT overlayer as examples we show that the exchange splitting one of the Dirac cones is effectively eliminated, giving rise to a half-quantized plateau when four non-magnetic blocks are interfaced with the MBT layer. Work-function alignment is of particular importance, as it dictates the resulting electronic spectrum and, consequently, the energy window over which the half-quantized plateau persists. The versatility of our approach offers a straightforward roadmap for future studies, both experimental and theoretical.
Overall, our findings provide the basis for developing strategies for engineering heterostructures and dopant profiles, not only to achieve the quantum anomalous and half-quantized anomalous Hall phases but also to facilitate switching between them.

\section{Calculation details}

The calculations were performed using the projector augmented-wave (PAW) method\cite{blochl1994projectoraugmentedwave}, as implemented in the VASP code\cite{kresse1993initiomolecular, kresse1996efficientiterative,kresse1996efficiencyabinitio}. Exchange-correlation effects were treated within the generalized gradient approximation (Perdew-Burke-Ernzerhof (PBE)  functional)\cite{perdew1996generalizedgradient}, with van der Waals interactions accounted of via the DFT-D3 approach\cite{grimme2010consistentaccurate}.

For Mn 3d states, we employed the GGA+U method~\cite{anisimov1991bandtheory} within the Dudarev scheme\cite{dudarev1998electronenergylossspectra}, where the Hubbard U parameters (5.3~eV for Mn) were determined using the linear-response method\cite{cococcioni2005linearresponse}.

The heterostructure was modeled using a periodic slab geometry with 12~\AA\ vacuum spacing. The work functions of the magnetic seven-layer films were calculated using the in-plane lattice parameters of the heterostructures. Atomic-type orbitals and corresponded tight-binding model were constructed via Wannier90~\cite{Pizzi_2020}. The calculations of the Hall conductivity have been done with with using the methodology described in Ref~\cite{varnava2018surfacesaxion}.

\begin{acknowledgments}
We gratefully thank V.N.~Men'shov for fruitful discussions. 
This work is supported by Saint-Petersburg State University for a research project 125022702939-2. 

\end{acknowledgments}

\bibliography{bib.bib}


\end{document}